\documentclass[longauth]{aa}

\usepackage{txfonts}
\usepackage{graphicx}
\usepackage{mathtools}
\usepackage{float}
\usepackage{epstopdf}
\usepackage{amsmath}

\bibpunct{(}{)}{;}{a}{}{,} % to follow the A&A style

\begin{document}

\title{LOFAR tied-array imaging and spectroscopy of solar S bursts}

%----------------- Authors ---------------

\author{D. E. Morosan\inst{\ref{1}}
\and 
P. T. Gallagher\inst{\ref{1}} 
 \and 
P. Zucca\inst{\ref{1}} 
 \and 
A. O'Flannagain\inst{\ref{1}} 
 \and 
R. Fallows\inst{\ref{2}} 
 \and 
H. Reid\inst{\ref{3}} 
 \and 
J. Magdaleni\'c\inst{\ref{4}} 
 \and 
G. Mann\inst{\ref{5}} 
 \and 
M. M. Bisi\inst{\ref{6}} 
 \and 
A. Kerdraon\inst{\ref{7}} 
 \and 
A. A. Konovalenko\inst{\ref{8}} 
 \and 
A. L. MacKinnon\inst{\ref{3}} 
 \and 
H. O. Rucker\inst{\ref{9}} 
 \and 
B. Thid\'e\inst{\ref{10}} 
 \and 
C. Vocks\inst{\ref{5}} 
 \and 
A.~Alexov\inst{\ref{11}} 
 \and 
J.~Anderson\inst{\ref{12}} 
 \and 
A.~Asgekar\inst{\ref{2}} \and \inst{\ref{13}} 
 \and 
I.~M.~Avruch\inst{\ref{14}} \and \inst{\ref{15}} 
 \and 
M.~J.~Bentum\inst{\ref{2}} \and \inst{\ref{16}} 
 \and 
G.~Bernardi\inst{\ref{17}} 
 \and 
A.~Bonafede\inst{\ref{18}} 
 \and 
F.~Breitling\inst{\ref{5}} 
 \and 
J.~W.~Broderick\inst{\ref{19}} \and \inst{\ref{20}} 
 \and 
W.~N.~Brouw\inst{\ref{2}} \and \inst{\ref{15}} 
 \and 
H.~R.~Butcher\inst{\ref{21}} 
 \and 
B.~Ciardi\inst{\ref{22}} 
 \and 
E.~de Geus\inst{\ref{2}} 
\and \inst{\ref{23}} 
 \and 
J.~Eisl\"offel\inst{\ref{24}} 
 \and 
H.~Falcke\inst{\ref{25}} 
\and \inst{\ref{2}} 
 \and 
W.~Frieswijk\inst{\ref{2}} 
 \and 
M.~A.~Garrett\inst{\ref{2}} 
\and \inst{\ref{26}} 
 \and 
J.~Grie\ss{}meier\inst{\ref{27}} 
\and \inst{\ref{28}} 
 \and 
A.~W.~Gunst\inst{\ref{2}} 
 \and 
J.~W.~T.~Hessels\inst{\ref{2}} 
\and \inst{\ref{29}} 
 \and 
M.~Hoeft\inst{\ref{24}} 
 \and 
A. ~Karastergiou\inst{\ref{19}} 
 \and 
V.~I.~Kondratiev\inst{\ref{2}} 
\and \inst{\ref{30}} 
 \and 
G.~Kuper\inst{\ref{2}} 
 \and 
J.~van Leeuwen\inst{\ref{2}} 
\and \inst{\ref{29}} 
 \and 
D.~McKay-Bukowski\inst{\ref{31}} 
\and \inst{\ref{32}} 
 \and 
J.~P.~McKean\inst{\ref{2}} 
\and \inst{\ref{15}} 
 \and 
H.~Munk\inst{\ref{2}} 
 \and 
E.~Orru\inst{\ref{2}} 
 \and 
H.~Paas\inst{\ref{33}} 
 \and 
R.~Pizzo\inst{\ref{2}} 
 \and 
A.~G.~Polatidis\inst{\ref{2}} 
 \and 
A.~M.~M.~Scaife\inst{\ref{20}} 
 \and 
J.~Sluman\inst{\ref{2}} 
 \and 
C.~Tasse\inst{\ref{7}} 
 \and 
M.~C.~Toribio\inst{\ref{2}} 
 \and 
R.~Vermeulen\inst{\ref{2}} 
 \and 
 P.~Zarka\inst{\ref{7}}
}
%----------------- Institutes ---------------
\institute{ School of Physics, Trinity College Dublin, Dublin 2, Ireland \label{1}
\and
ASTRON, Netherlands Institute for Radio Astronomy, Postbus 2, 7990 AA Dwingeloo, The Netherlands \label{2}
\and
School of Physics and Astronomy, SUPA, University of Glasgow, Glasgow G12 8QQ, United Kingdom \label{3}
\and
Solar-Terrestrial Center of Excellence, SIDC, Royal Observatory of Belgium, Avenue Circulaire 3, B-1180 Brussels, Belgium \label{4}
\and
Leibniz-Institut f\"{u}r Astrophysik Potsdam (AIP), An der Sternwarte 16, 14482 Potsdam, Germany \label{5}
\and
RAL Space, Science and Technology Facilities Council, Rutherford Appleton Laboratory, Harwell Oxford, Oxfordshire, OX11 OQX, United Kingdom \label{6}
\and
LESIA, UMR CNRS 8109, Observatoire de Paris, 92195   Meudon, France \label{7}
\and
Institute of Radio Astronomy, 4, Chervonopraporna Str., 61002 Kharkiv, Ukraine \label{8}
\and
Commission for Astronomy, Austrian Academy of Sciences, Schmiedlstrasse 6, 8042 Graz, Austria \label{9}
\and
Swedish Institute of Space Physics, Box 537, SE-75121 Uppsala, Sweden \label{10}
\and
Space Telescope Science Institute, 3700 San Martin Drive, Baltimore, MD 21218, USA \label{11}
\and
Helmholtz-Zentrum Potsdam, DeutschesGeoForschungsZentrum GFZ, Department 1: Geodesy and Remote Sensing, Telegrafenberg, A17, 14473 Potsdam, Germany \label{12}
\and
Shell Technology Center, Bangalore, India \label{13}
\and
SRON Netherlands Insitute for Space Research, PO Box 800, 9700 AV Groningen, The Netherlands \label{14}
\and
Kapteyn Astronomical Institute, PO Box 800, 9700 AV Groningen, The Netherlands \label{15}
\and
University of Twente, The Netherlands \label{16}
\and
Harvard-Smithsonian Center for Astrophysics, 60 Garden Street, Cambridge, MA 02138, USA \label{17}
\and
University of Hamburg, Gojenbergsweg 112, 21029 Hamburg, Germany \label{18}
\and
Astrophysics, University of Oxford, Denys Wilkinson Building, Keble Road, Oxford OX1 3RH, UK \label{19}
\and
School of Physics and Astronomy, University of Southampton, Southampton, SO17 1BJ, UK \label{20}
\and
Research School of Astronomy and Astrophysics, Australian National University, Mt Stromlo Obs., via Cotter Road, Weston, A.C.T. 2611, Australia \label{21}
\and
Max Planck Institute for Astrophysics, Karl Schwarzschild Str. 1, 85741 Garching, Germany \label{22}
\and
SmarterVision BV, Oostersingel 5, 9401 JX Assen, The Neterlands \label{23}
\and
Th\"{u}ringer Landessternwarte, Sternwarte 5, D-07778 Tautenburg, Germany \label{24}
\and
Department of Astrophysics/IMAPP, Radboud University Nijmegen, P.O. Box 9010, 6500 GL Nijmegen, The Netherlands \label{25}
\and
Leiden Observatory, Leiden University, PO Box 9513, 2300 RA Leiden, The Netherlands \label{26}
\and
LPC2E - Universite d'Orleans/CNRS, 3A, Avenue de la Recherche Scientifique, 45071, Orl\'{e}ans cedex 2, France \label{27}
\and
Station de Radioastronomie de Nancay, Observatoire de Paris - CNRS/INSU, USR 704 - Univ. Orleans, OSUC , route de Souesmes, 18330 Nancay, France \label{28}
\and
Anton Pannekoek Institute, University of Amsterdam, Postbus 94249, 1090 GE Amsterdam, The Netherlands \label{29}
\and
Astro Space Center of the Lebedev Physical Institute, Profsoyuznaya str. 84/32, Moscow 117997, Russia \label{30}
\and
Sodankyl\"{a} Geophysical Observatory, University of Oulu, T\"{a}htel\"{a}ntie 62, 99600 Sodankyl\"{a}, Finland \label{31}
\and
STFC Rutherford Appleton Laboratory,  Harwell Science and Innovation Campus,  Didcot  OX11 0QX, UK \label{32}
\and
Center for Information Technology (CIT), University of Groningen, The Netherlands \label{33}
}

\date{ Received /
		Accepted }

\abstract{The Sun is an active source of radio emission that is often associated with energetic phenomena ranging from nanoflares to coronal mass ejections (CMEs). At low radio frequencies ($<$100~MHz), numerous millisecond duration radio bursts have been reported, such as radio spikes or solar S bursts (where S stands for short). To date, these have neither been studied extensively nor imaged because of the instrumental limitations of previous radio telescopes. }
{Here, Low Frequency Array (LOFAR) observations were used to study the spectral and spatial characteristics of a multitude of S bursts, as well as their origin and possible emission mechanisms.}
{We used 170 simultaneous tied-array beams for spectroscopy and imaging of S bursts. Since S bursts have short timescales and fine frequency structures, high cadence ($\sim$50~ms) tied-array images were used instead of standard interferometric imaging, that is currently limited to one image per second.}
{On 9 July 2013, over 3000 S bursts were observed over a time period of $\sim$8 hours. S bursts were found to appear as groups of short-lived ($<$1~s) and narrow-bandwidth ($\sim$2.5~MHz) features, the majority drifting at $\sim$3.5~MHz s$^{-1}$ and a wide range of circular polarisation degrees (2--8 times more polarised than the accompanying Type III bursts). Extrapolation of the photospheric magnetic field using the potential field source surface (PFSS) model suggests that S bursts are associated with a trans-equatorial loop system that connects an active region in the southern hemisphere to a bipolar region of plage in the northern hemisphere.}
{We have identified polarised, short-lived solar radio bursts that have never been imaged before. They are observed at a height and frequency range where plasma emission is the dominant emission mechanism, however they possess some of the characteristics of electron-cyclotron maser emission.  }

\keywords{Sun: corona -- Sun: radio radiation -- Sun: particle emission -- Sun: magnetic fields}

\maketitle

\section{Introduction}

{The Sun is an active star that produces numerous large scale energetic phenomena often accompanied by solar radio bursts such as the Type I, II, III, IV and V radio bursts. At decimetre/metre wavelengths, numerous other radio bursts have been observed, for example metric spikes \citep{be82, be96}, drifting spikes \citep{el79} and supershort solar radio bursts \citep{ma06}. These bursts exhibit a variety of fine frequency structures and time scales of $<$1~s, which can be an indicator of small scale processes occurring in the solar corona. However, at longer wavelengths, corresponding to frequencies $<$100~MHz, there have been very few studies of short fine structure radio bursts due to sensitivity and imaging limitations of previous radio telescopes. }

\begin{figure*}[ht]
\includegraphics[width = 250 px, angle = -90, trim = 140px 50px 100px 20px ]{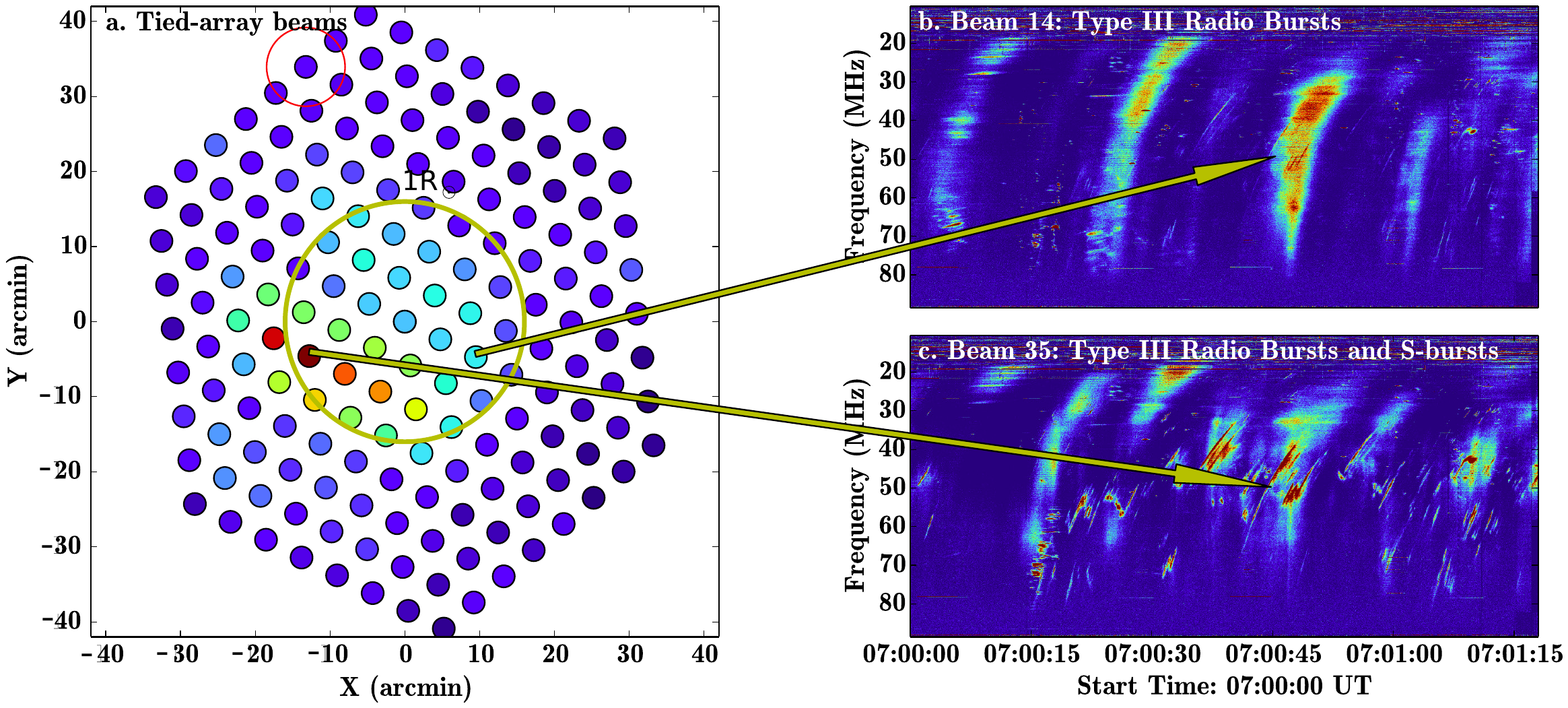}
\caption{LOFAR tied-array beam observations of Type III radio bursts and solar S bursts. (a) Map of 170 tied-array beams covering a field-of-view of $\sim$1.3\degr about the Sun. The full-width half-maximum (FWHM) of the beams at zenith at a frequency of 60~MHz is represented by the red circle and the size of the optical Sun is represented by the yellow circle. (b) Dynamic spectrum recorded for a period of 1.3 minutes corresponding to Beam 14 containing Type III radio bursts. (c) Dynamic spectrum recorded for a period of 1.3 minutes corresponding to Beam 35 containing Type III radio bursts and S bursts. The two arrows indicate the beams that recorded the two dynamic spectra in (b) and (c) pointing at the time and frequency corresponding to the intensity values in (a).\newline \label{fig1}}
\end{figure*}

\begin{figure}[t]
\includegraphics[width = 285px, trim = 70px 20px 0px 60px ]{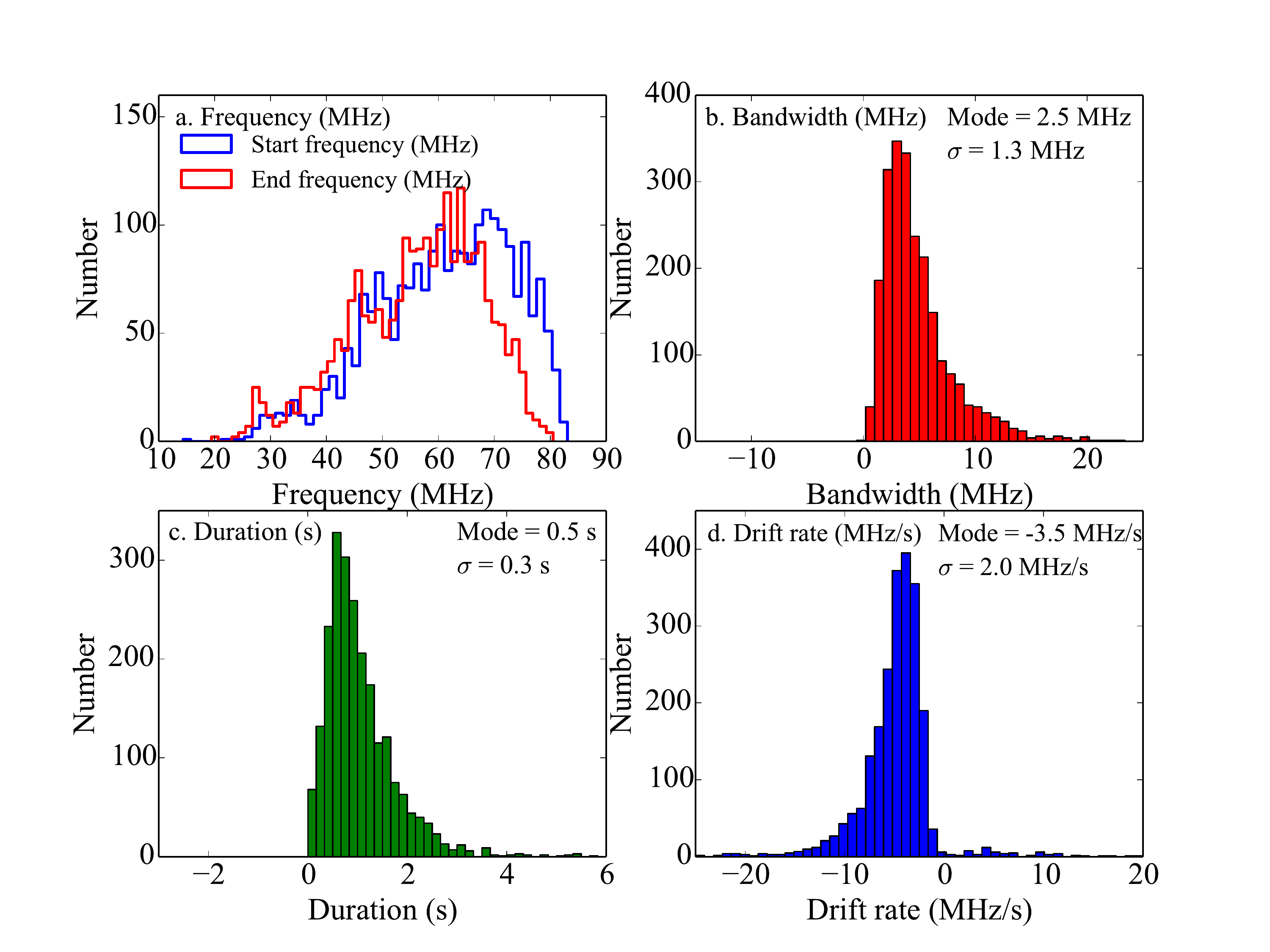}
\caption{S-burst property distributions: start and end frequency (a), total frequency bandwidth (b), duration (c) and drift rate (d). The most frequently occurring value (mode) and the standard deviation are shown. The standard deviation was calculated using a normal distribution fit to each of the distributions.\newline \label{fig2}}
\end{figure}

{At low frequencies, below 100~MHz, \citet{el69} used 32 broad-band dipoles to identify a new type of short duration radio burst that he named fast drift storm bursts. Later, \citet{mcc82} observed these bursts using the Llanherne radio telescope consisting of 4096 dipoles \citep{fe80}. He showed that their drift rate is slower than that of Type III radio bursts (about 1/3 the drift rate of Type IIIs). \citet{mcc82} renamed fast drift storms as ``solar S bursts" due to their similarity to jovian S bursts. S bursts appear as narrow drifting lines in dynamic spectra with a temporal width of up to a few tens of milliseconds and total duration of about 1~s. They have been observed in a frequency range of 30--150~MHz \citep{el82, mcc82}. They have short instantaneous bandwidths of about 120~kHz and instantaneous duration of 50~ms at frequencies of 40~MHz \citep{mcc83}. More recently, \citet{bri08} and \citet{mel10} have reported observations of S bursts in the frequency range 10--30~MHz. These were observed with the UTR-2 radio telescope consisting of over 2000 dipoles with a total area of over 30000 m$^2$ \citep{abr01, mel11}. Most S bursts have negative drift rates, i.e. from high to low frequencies, however a few S bursts have positive drift rates \citep{mcc82, bri08}. All S bursts occurred during times of other solar activity such as Type III and Type IIIb radio bursts \citep{mcc82, bri08, mel10}. The majority of S burst studies consider plasma emission as the most probable emission mechanism \citep{ze86, mel10}. To date, S bursts have only been observed with highly sensitive instruments as they have a low intensity compared to the majority of radio bursts. So far, there has been no imaging of S bursts for more detailed studies of their origins and emission mechanisms.}

\begin{table}[t]
\caption{\label{table1}S bursts characteristics estimated from LOFAR dynamic spectra are compared to results by \citet{mcc82}.}
\centering
	\begin{tabular}{lcc} 
	\hline\hline
	 & LOFAR & McConnell\\ \hline \hline
	Frequency range (MHz) & 20--80 & 30--82 \\ 
	Duration (s) & 0.5$\pm$0.3 & 1.2\\ 
	Instantaneous duration (ms) & 20--400 & 50 \\ 
	Total bandwidth (MHz) & 2.5$\pm$1.3 & $<$5 \\ 
	Instantaneous bandwidth (MHz) & 0.1--1.5 & 0.1\\ 
	Drift rate (MHz s$^{-1}$) & -3.5$\pm$2.0  & -3 \\ \hline \hline
	\end{tabular}
\end{table}

{With the recent development of the Low Frequency Array \citep[LOFAR,][]{lofar13}, we now have the possibility to observe solar S bursts in high time and frequency resolution dynamic spectra and for the first time obtain spatial characteristics of S bursts from imaging. In this paper, LOFAR tied-array beams were used to study the spatial and spectral characteristics of S bursts. In Section 2 we give an overview of the LOFAR instrument and the observational method. In Section 3 we present the results of the spectral and imagining analysis which are discussed in Section 4.\newline}

\begin{figure*}[ht]
\centering
\includegraphics[width = 420px,  trim = 0px 60px 0px 15px]{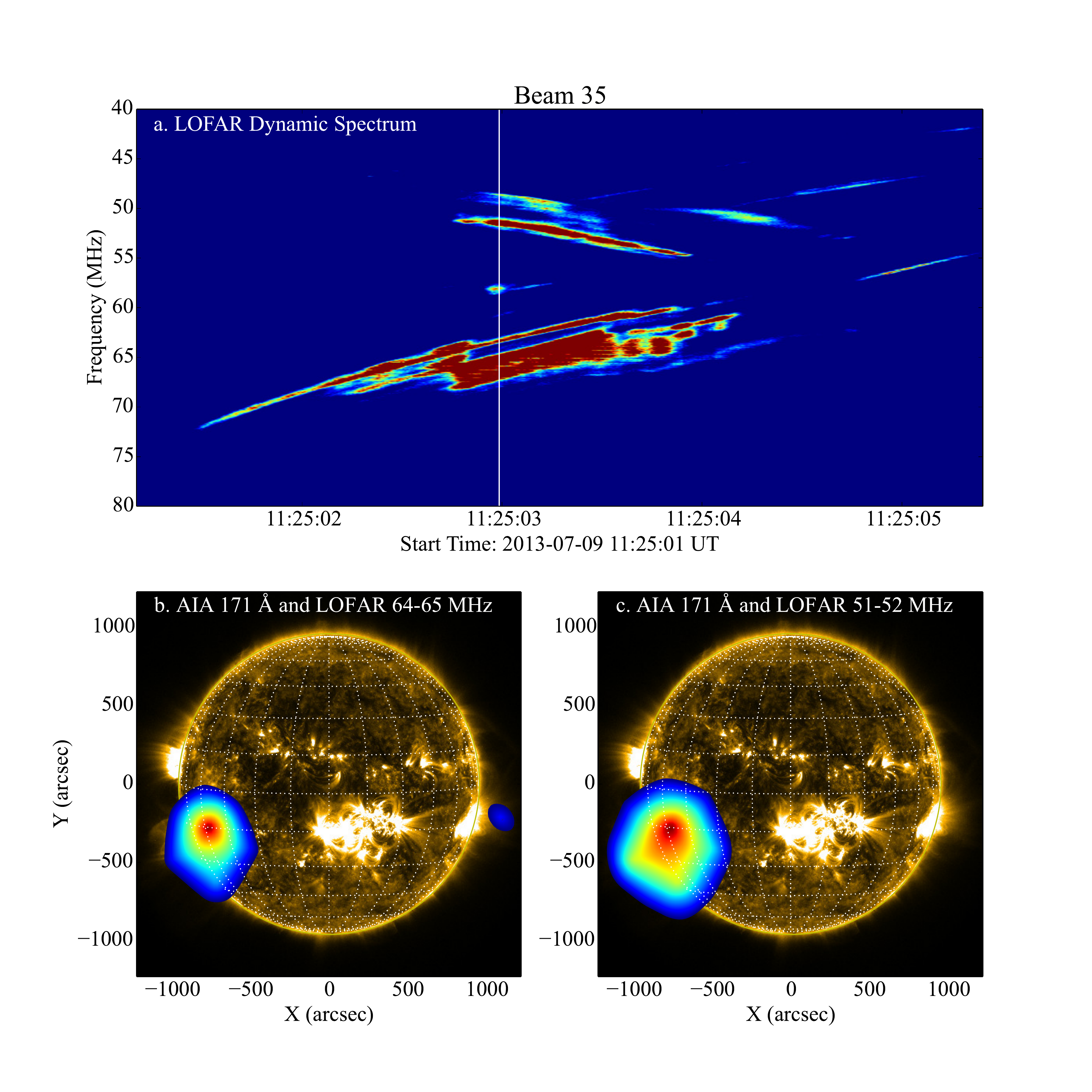}
\caption{Location of S bursts at two separate frequencies. (a) LOFAR 4-second dynamic spectrum showing negatively and positively drifting S bursts. The white line marks the timestamp in the images below. (b) SDO/AIA 171 \AA~image at 11:25:11 UT and the overlaid LOFAR source of the negative drifting S burst at a frequency of 64--65~MHz at the time marked by the white line in (a). (c) SDO/AIA 171 \AA~image 11:25:11 UT and the overlaid LOFAR source of the positive drifting S bursts at a frequency of 51--52~MHz at the time marked by the white line in (a). An animation of the temporal evolution is available in the online edition.\label{fig3}}
\end{figure*}

\section{Observations and data analysis}

{On 9 July 2013 starting at 07:00 UT, over 3000 S bursts were observed during a LOFAR observation campaign that lasted until 14:30 UT. These observations were made at times of medium solar activity in the form of three C-class flares and the presence of four $\beta\gamma\delta$ active regions (NOAA 11784, 11785, 11787, 11789) on the visible side of the solar disc. In this time interval, a long-lived noise storm was observed above active region NOAA 11785 at frequencies greater than 150~MHz. }

{LOFAR represents a new milestone in low radio frequency arrays and was constructed by the Netherlands Institute for Radio Astronomy (ASTRON). It consists of $\sim$7000 antennas: Low Band Antennas (LBAs) operating at frequencies of 10--90~MHz and High Band Antennas (HBAs) which operate at 110--240~MHz \citep{lofar13}. These antennas are distributed in 24 core stations and 14 remote stations across the Netherlands and 9 international stations across Europe. }

{In this paper, we used one of LOFAR's beam formed modes \citep{sta11, lofar13} in the LBA frequency range. Figure 1a shows 170 simultaneous beams which were used to observe the Sun covering a field-of-view of 1.3\degr~centred on the Sun. The 24 LOFAR core stations were used to produce these beams. Each beam produces a high time and frequency resolution dynamic spectra ($\sim$10~ms; 12.5~kHz) as can be seen in Figures 1b and 1c, in which the radio bursts can be identified. Since each beam has a spatial location, it can be used to extract the intensity of the radiation at that specific location. These intensity values can be plotted as `macro-pixels' onto the tied array map for a chosen time and frequency as seen in Figure 1a. The arrows in Figure 1 denote the time, frequency and beam where the intensity value was extracted. }

{Radio sources can be imaged by interpolating between the intensity of radiation from all beams and producing tied-array images of radio bursts \citep[for more details, see][]{mo14}.  All images of S bursts were averaged over a frequency bin of 1~MHz to reduce spectral noise and over a time period of 50~ms. Side lobes are often visible in tied-array images due to the coherent combination of tied-array beams, however, most side lobes have been excluded manually as they occur at the same time as the source and always at the same location some distance away from the source. Due to the size of tied-array beams there is an uncertainty in determining the exact location of the radio source as the source could be located anywhere within the beam. This uncertainty is estimated to be 0.2~$R_\sun$ at 80~MHz and 0.6~$R_\sun$ at 30~MHz and it corresponds to the half power beam width. The main source of uncertainty in beam pointing is the ionosphere, which is most pronounced during periods of elevated solar activity. During our observations, there was no significant X-ray activity other than a few C-class flares. In addition, the centroid position of our sources was found not to vary significantly over many hours. }   

\begin{figure}[t]
\includegraphics[angle = 0, width = 300px, trim =180px 80px 100px 55px]{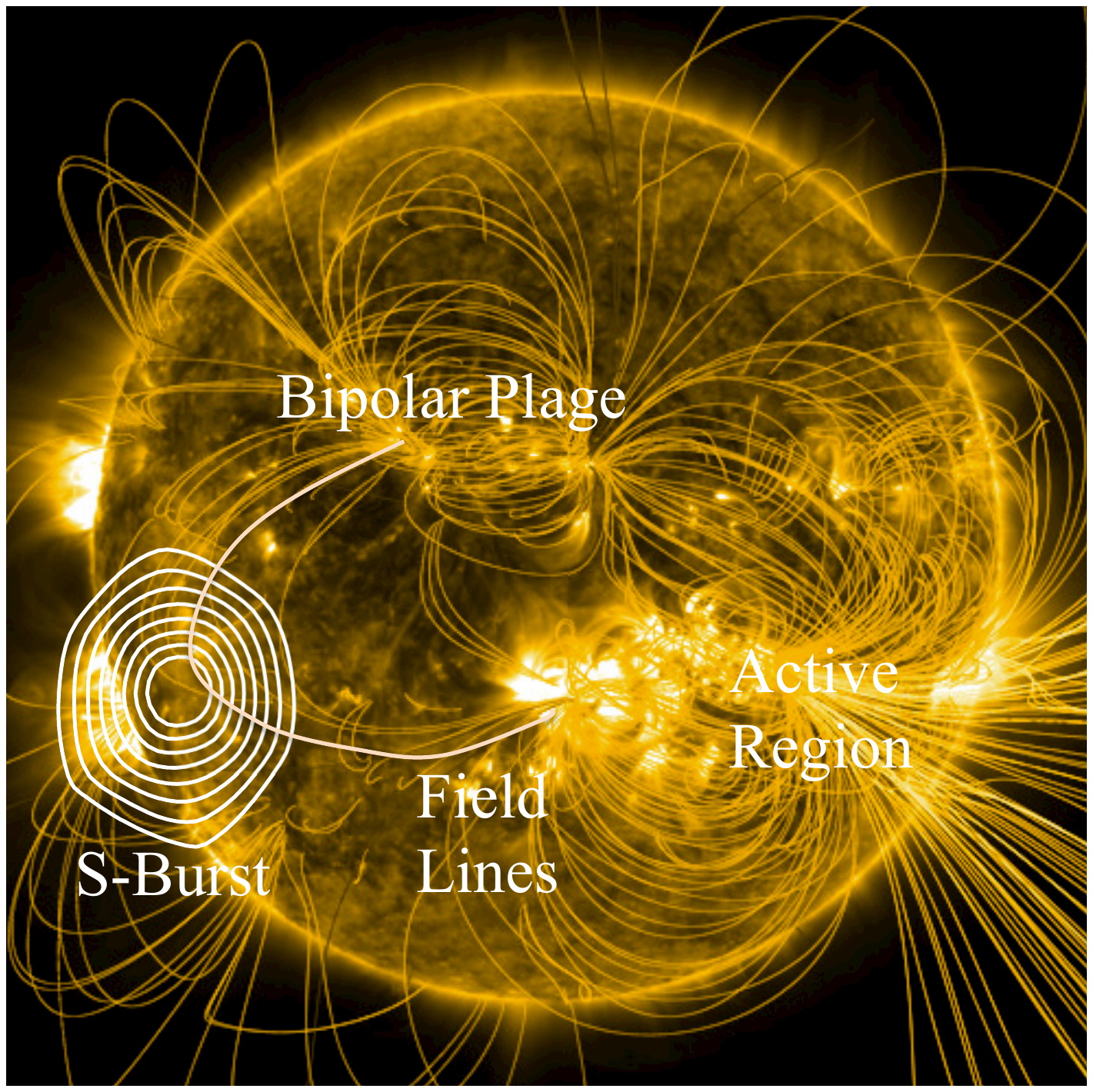}
\caption{S bursts source position and their relation to the coronal magnetic field. The AIA 171~\AA~ image at 11:24:12 UT is overlaid PFSS magnetic field extrapolations. The white contours represent the LOFAR S bursts source at 11:24:56 UT and a frequency of 74--75~MHz. One of the trans-equatorial loops connecting the bipolar page to the active region is labelled in the figure. \label{fig4}}
\end{figure}

\section{Results}

\subsection{S bursts spectral characteristics} 

{An example of S bursts that occurred during a time period of 1.3 minutes on 9 July 2013 at 07:00 UT is shown in Figure 1c. The S bursts were recorded simultaneously with other solar activity mainly in the form of Type III and Type IIIb radio bursts, some of which did not originate at the same location as the S bursts as can be seen in Figures 1b and 1c. }

{Over 3000 S bursts have been recorded during a time interval of 8 hours on the same day in the LBA frequency range. Due to a variety of shapes (various lengths, time profiles, frequency profiles) and various intensity levels, similar to Type III/Type IIIb bursts, it was necessary to identify these bursts manually. Distributions of their characteristics such as the start and end frequency, total bandwidth, duration and drift rate are shown in Figure 2. These distributions are asymmetrical and therefore the mode of the distributions is shown instead of the mean due to the fact that they are skewed. Most S bursts had short durations of $<$1~s and instantaneous durations at a fixed frequency of $<$400~ms, some as small as 20~ms. They were observed at frequencies between 20 to 80~MHz and the majority had a total bandwidth of $\sim$2.5$\pm$1.3~MHz and instantaneous bandwidths at a fixed time varying between 0.1--1.5~MHz. Very few S bursts had longer bandwidths $>$20~MHz. Their drift rates varied with frequency from -1~MHz~s$^{-1}$ at frequencies of 20~MHz to -7~MHz~s$^{-1}$ at 75~MHz (Figure 2) but most bursts had a drift rate of -3.5$\pm$2.0~MHz s$^{-1}$. The standard deviation of all distributions is a significant fraction of the mode, therefore there is a wide spread in duration, bandwidth and drift rate of S bursts.}

{Table 1 shows that these values are comparable to the findings of \citet{mcc82, mcc83}. A few S bursts were found to have positive drift rates and an example is shown in Figure 3a. Positive drifting S bursts have also been reported before by \citet{mcc82, mcc83, bri08, mel10}.}

{Full Stokes data were recorded starting at 11:40:00 UT for a duration of 1~hour. During this time several hundred S bursts were observed as well as Type III radio bursts. The observed data is uncalibrated and we cannot estimate the degree of polarisation accurately, however, we can compare the S bursts polarisation to the background Type III radio bursts. Type IIIs emitted at the harmonic of the local plasma frequency are believed to be weakly polarised \citep[$<$10\%,][]{du80}. We have found a wide range of different circular polarisation degrees for S bursts from 2--8 times more polarised than the accompanying Type III bursts. This agrees with previous studies in which the S bursts studied were found to have a strong component of circular polarisation \citep{mcc82}.}

\subsection{S bursts spatial characteristics} 

{An example of a number of S bursts analysed in more detail is shown in Figure 3. Figure 3 is a snapshot from the movie accompanying this paper. Figure 3a shows a 4~second LOFAR dynamic spectrum containing S bursts with both negative and positive drift rates. Figures 3b and 3c show the location of these S bursts overlaid on extreme ultraviolet images (EUV) at 171~\AA~from the Atmospheric Imaging Assembly \citep[AIA;][]{le12} onboard NASA's Solar Dynamic Observatory (SDO). S bursts are imaged at frequencies of 64--65~MHz to sample the negatively drifting S bursts and 51--52~MHz to sample the positively drifting S bursts. }

{Tied-array imaging can produce images at very high cadence (50~ms) which is necessary for radio bursts with durations of $<$1~s. However the beams have a FWHM of 12.6\arcmin~at a frequency of 50~MHz as they can only be produced by the LOFAR core. Its 2~km baseline is too small to produce arcsecond beam sizes. As a result the source sizes appear large. We cannot detect any movement in the source centroids due to frequency change. Both negatively and positively drifting S bursts appear to originate at the same location. All S bursts analysed in this observation campaign on 9 July 2013 appear to originate at the same location with insignificant deviations in the source centroids ($<$$0.1~R_{\odot}$). It is possible that this results from the reduced spatial resolution of tied-array beams imaging. }

{We investigated the origin of these S bursts by relating the source locations to the coronal magnetic field. Figure 4 shows that the radio source does not seem to be directly located on top of known active regions but it appears to be close to the top of trans-equatorial loops connecting a complex active region in the southern hemisphere and a large area of bipolar plage in the northern hemisphere. The top of these equatorial loops is located at a distance of about $\sim$1.8~$R_\odot$ from the solar centre. This distance is calculated from the 3D PFSS extrapolation of the loop and it represents the radial height of the loop top from the surface. The spatial coincidence of the S bursts and the top of these equatorial loops in Figure 4 suggests that S bursts are associated with these loops.}

\section{Discussion and conclusion}
 
{Over 3000 S bursts were observed by LOFAR at times of medium solar activity and simultaneously with other radio emission such as Type III and Type IIIb bursts. Most S bursts had very short durations ($\sim$0.5~s) and short bandwidths ($\sim$2.5~MHz). Their drift rates varied with frequency with the majority drifting at -3.5~MHz s$^{-1}$ at frequencies of 20--80~MHz (Table 1) which is about 1/3 of the Type III drift rate. Type IIIs drift at -11.33~MHz s$^{-1}$ in the frequency range of 40--70~MHz according to \citet{mann02}. The majority of S bursts are found at frequencies between 60--70~MHz (Figure 2a), but this result may be influenced by the 60~MHz peak in the sensitivity of LBAs antennas \citep{lofar13}. S bursts were found to be significantly more polarised than Type IIIs. Calibrated polarisation observations of S bursts will allow for these properties to be further studied in a follow on paper. }

{For the first time, we were able to image solar S bursts. Tied-array imaging shows that all S bursts originate in the same spatial location with very little deviation of the source centroid. There seem to be no features in the solar corona at the location of S bursts sources other than long trans-equatorial loops connecting an active region in the southern hemisphere and an extended area of bipolar plage in the northern hemisphere. We have investigated two possible emission mechanisms responsible for solar S bursts: plasma emission and electron-cyclotron maser (ECM) emission which are discussed below.}

{So far, plasma emission has been proposed by \citet{ze86} and \citet{mel10} as the mechanism responsible for emission of S bursts. \citet{ze86} suggest that electrons with velocities 10--20 times above their thermal velocity excite plasma waves near the upper hybrid resonance frequency. These waves are then scattered by ions producing electromagnetic waves at the plasma frequency (\textit{f$_p$}). \citet{mel10} propose a model in which S bursts are generated by the coalescence of magnetosonic waves and Langmuir waves and derive a minimum magnetic field of 2~G necessary for the generation of these bursts at the plasma frequency (\textit{f$_p$}). }

{In our observation, S bursts appear to be associated with equatorial loops stretching to high altitudes ($\sim$1.8~$R_\odot$) where plasma emission is likely the dominant emission mechanism \citep{ga89}. Knowing the height of these loops, a density model can be used to find the electron density at this distance. Using the electron density we can estimate the frequency of plasma radiation at the top of the loops using the following equation:
\begin{align}
\label{eq:1}
f = n f_p = n~C\sqrt{N_e} .
\end{align} 
The emission frequency, $f$, is given by the local plasma frequency, $f_p$, multiplied by the harmonic number $n$ and is directly proportional to the square root of the electron density, $N_e$, in cm$^{-3}$, where $C = 8980$~Hz~cm$^{3/2}$ is the constant of proportionality. }

{We used the radial electron density models of \citet{new61}, \citet{sa77} and \citet{mann99} and the time-dependent density model of \citet{zu14} to estimate electron densities at specific distances. These models predict electron densities in the range of 1.2$\times10^6$--1.1$\times10^7$~cm$^{-3}$ at $\sim$1.8~$R_\odot$. This corresponds to a frequency range of $\sim$10--30~MHz for fundamental plasma emission and $\sim$20--60~MHz for harmonic plasma emission, frequency ranges in which S bursts are found. In addition, S bursts show a drift rate dependence with frequency similar to that of Type III radio bursts \citet{mcc80}. A similar drift rate dependence with frequency was also found in our analysis which will be discussed in a follow-on paper. It is therefore possible that S bursts are a form of plasma emission. The high degree of polarisation of some of these bursts requires that they are emitted at the fundamental plasma frequency \citep{we84}.  }

{However, S bursts have a significantly slower drift rate than Type III radio bursts (about 1/3 of the Type III drift rate), very narrow bandwidths and short lifetimes and they are significantly more polarised. This is indicative of ECM emission. Jovian S bursts, which are similar in appearance to solar S bursts, originate due to cyclotron maser emission in the flux tubes connecting Io or Io's wake to Jupiter \citep{he07}. For ECM to occur in the solar corona, the electron-cyclotron frequency ($f_B = eB/2\pi m_e$, where $B$ is the magnetic field) has to be greater than the plasma frequency \citep{mel91}. Therefore we require a high $B$ field at the emission site in the corona ($>20$~G), as well as low electron plasma density ($N_e$). While the plasma density is low at large heights in the corona where S bursts occur, the $B$ field is small ($\sim$0.25~G) as estimated from the models of \citet{zu14}. From our analysis, the location of these S bursts does not meet the condition for ECM to occur. Presently we are not able to draw a conclusion on the emission mechanism of S bursts and we hope that future imaging observations of S bursts will constrain these results.}

{The tied-array beams prove promising in detecting millisecond duration radio bursts with very high frequency and temporal resolution. In addition to this, tied-array beam observations can show the positions of these radio bursts sources. The accuracy in these positions is only limited by the beam size. Ionospheric effects were not significant during this observation. No movement was detected in the S bursts source position and the Quiet Sun position was constant during the observation. The spatial resolution of tied-array beams can only be increased by increasing the baseline of the LOFAR beam-formed modes (i.e., extend beyond the LOFAR core) which would not have sufficient beams to cover the Sun and is also currently not available as an observational mode. However, at such low frequencies, scattering in the corona may limit the resolution of the beam sizes. Another way to increase the spatial resolution when observing S bursts is to decrease the cadence of LOFAR's interferometric snapshot imaging to $<$0.1~s which is also currently not ideal due to the high data volume rates. This, however, can be compensated by a coarser frequency resolution and it is worth testing in the future. Tied-array beam analysis has been previously applied to Type III radio bursts \citep{mo14} and it shows a high potential for studying S bursts and other radio bursts within the LOFAR spectral range.}

\begin{acknowledgements}{This work has been supported by a Government of Ireland studentship from the Irish Research Council (IRC), the Non-Foundation Scholarship awarded by Trinity College Dublin, the Innovation Academy and the IRC New Foundations. Hamish Reid is supported by a SUPA Advanced Fellowship and an STFC grant ST/L000741/1. We would finally like to acknowledge the LOFAR telescope. LOFAR, the Low Frequency Array designed and constructed by ASTRON, has facilities in several countries, that are owned by various parties (each with their own funding sources), and that are collectively operated by the International LOFAR Telescope (ILT) foundation under a joint scientific policy. }
\end{acknowledgements}

% for the bibliography, at the end
\bibliographystyle{aa} % style aa.bst
\bibliography{Morosan_paper_AA} % your references file.bib

\begin{thebibliography}{29}
\expandafter\ifx\csname natexlab\endcsname\relax\def\natexlab#1{#1}\fi

\bibitem[{{Abranin} {et~al.}(2001){Abranin}, {Bruck}, {Zakharenko}, \&
  {Konovalenko}}]{abr01}
{Abranin}, E.~P., {Bruck}, Y.~M., {Zakharenko}, V.~V., \& {Konovalenko}, A.~A.
  2001, Experimental Astronomy, 11, 85

\bibitem[{{Benz} {et~al.}(1996){Benz}, {Csillaghy}, \& {Aschwanden}}]{be96}
{Benz}, A.~O., {Csillaghy}, A., \& {Aschwanden}, M.~J. 1996, \aap, 309, 291

\bibitem[{{Benz} {et~al.}(1982){Benz}, {Jaeggi}, \& {Zlobec}}]{be82}
{Benz}, A.~O., {Jaeggi}, M., \& {Zlobec}, P. 1982, \aap, 109, 305

\bibitem[{{Briand} {et~al.}(2008){Briand}, {Zaslavsky}, {Maksimovic}, {Zarka},
  {Lecacheux}, {Rucker}, {Konovalenko}, {Abranin}, {Dorovsky}, {Stanislavsky},
  \& {Melnik}}]{bri08}
{Briand}, C., {Zaslavsky}, A., {Maksimovic}, M., {et~al.} 2008, \aap, 490, 339

\bibitem[{{Dulk} \& {Suzuki}(1980)}]{du80}
{Dulk}, G.~A. \& {Suzuki}, S. 1980, \aap, 88, 203

\bibitem[{{Elgaroy} \& {Sveen}(1979)}]{el79}
{Elgaroy}, O. \& {Sveen}, O.~P. 1979, \nat, 278, 626

\bibitem[{{Ellis}(1969)}]{el69}
{Ellis}, G.~R.~A. 1969, Australian Journal of Physics, 22, 177

\bibitem[{{Ellis}(1982)}]{el82}
{Ellis}, G.~R.~A. 1982, Australian Journal of Physics, 35, 87

\bibitem[{{Ferris} {et~al.}(1980){Ferris}, {Turner}, {Hamilton}, \&
  {McCulloch}}]{fe80}
{Ferris}, R.~H., {Turner}, P.~J., {Hamilton}, P.~A., \& {McCulloch}, P.~M.
  1980, Proceedings of the Astronomical Society of Australia, 4, 26

\bibitem[{{Gary} \& {Hurford}(1989)}]{ga89}
{Gary}, D.~E. \& {Hurford}, G.~J. 1989, Washington DC American Geophysical
  Union Geophysical Monograph Series, 54, 237

\bibitem[{{Hess} {et~al.}(2007){Hess}, {Zarka}, \& {Mottez}}]{he07}
{Hess}, S., {Zarka}, P., \& {Mottez}, F. 2007, \planss, 55, 89

\bibitem[{{Lemen} {et~al.}(2012){Lemen}, {Title}, {Akin}, {Boerner}, {Chou},
  {Drake}, {Duncan}, {Edwards}, {Friedlaender}, {Heyman}, {Hurlburt}, {Katz},
  {Kushner}, {Levay}, {Lindgren}, {Mathur}, {McFeaters}, {Mitchell}, {Rehse},
  {Schrijver}, {Springer}, {Stern}, {Tarbell}, {Wuelser}, {Wolfson}, {Yanari},
  {Bookbinder}, {Cheimets}, {Caldwell}, {Deluca}, {Gates}, {Golub}, {Park},
  {Podgorski}, {Bush}, {Scherrer}, {Gummin}, {Smith}, {Auker}, {Jerram},
  {Pool}, {Soufli}, {Windt}, {Beardsley}, {Clapp}, {Lang}, \& {Waltham}}]{le12}
{Lemen}, J.~R., {Title}, A.~M., {Akin}, D.~J., {et~al.} 2012, \solphys, 275, 17

\bibitem[{{Magdaleni{\'c}} {et~al.}(2006){Magdaleni{\'c}}, {Vr{\v s}nak},
  {Zlobec}, {Hillaris}, \& {Messerotti}}]{ma06}
{Magdaleni{\'c}}, J., {Vr{\v s}nak}, B., {Zlobec}, P., {Hillaris}, A., \&
  {Messerotti}, M. 2006, \apjl, 642, L77

\bibitem[{{Mann} {et~al.}(1999){Mann}, {Jansen}, {MacDowall}, {Kaiser}, \&
  {Stone}}]{mann99}
{Mann}, G., {Jansen}, F., {MacDowall}, R.~J., {Kaiser}, M.~L., \& {Stone},
  R.~G. 1999, \aap, 348, 614

\bibitem[{{Mann} \& {Klassen}(2002)}]{mann02}
{Mann}, G. \& {Klassen}, A. 2002, in ESA Special Publication, Vol. 506, Solar
  Variability: From Core to Outer Frontiers, ed. A.~{Wilson}, 245--248

\bibitem[{{McConnell}(1980)}]{mcc80}
{McConnell}, D. 1980, Proceedings of the Astronomical Society of Australia, 4,
  64

\bibitem[{{McConnell}(1982)}]{mcc82}
{McConnell}, D. 1982, \solphys, 78, 253

\bibitem[{{McConnell}(1983)}]{mcc83}
{McConnell}, D. 1983, \solphys, 84, 361

\bibitem[{{Melnik} {et~al.}(2010){Melnik}, {Konovalenko}, {Rucker},
  {Dorovskyy}, {Abranin}, {Lecacheux}, \& {Lonskaya}}]{mel10}
{Melnik}, V.~N., {Konovalenko}, A.~A., {Rucker}, H.~O., {et~al.} 2010,
  \solphys, 264, 103

\bibitem[{{Melnik} {et~al.}(2011){Melnik}, {Rucker}, {Konovalenko},
  {Dorovskyy}, {Abranin}, \& {Lecacheux}}]{mel11}
{Melnik}, V.~N., {Rucker}, H.~O., {Konovalenko}, A.~A., {et~al.} 2011,
  Planetary, Solar and Heliospheric Radio Emissions (PRE VII), 343

\bibitem[{{Melrose}(1991)}]{mel91}
{Melrose}, D.~B. 1991, \araa, 29, 31

\bibitem[{{Morosan} {et~al.}(2014){Morosan}, {Gallagher}, {Zucca}, {Fallows},
  {Carley}, {Mann}, {Bisi}, {Kerdraon}, {Konovalenko}, {MacKinnon}, {Rucker},
  {Thid{\'e}}, {Magdaleni{\'c}}, {Vocks}, {Reid}, {Anderson}, {Asgekar},
  {Avruch}, {Bentum}, {Bernardi}, {Best}, {Bonafede}, {Bregman}, {Breitling},
  {Broderick}, {Br{\"u}ggen}, {Butcher}, {Ciardi}, {Conway}, {de Gasperin}, {de
  Geus}, {Deller}, {Duscha}, {Eisl{\"o}ffel}, {Engels}, {Falcke}, {Ferrari},
  {Frieswijk}, {Garrett}, {Grie{\ss}meier}, {Gunst}, {Hassall}, {Hessels},
  {Hoeft}, {H{\"o}randel}, {Horneffer}, {Iacobelli}, {Juette}, {Karastergiou},
  {Kondratiev}, {Kramer}, {Kuniyoshi}, {Kuper}, {Maat}, {Markoff}, {McKean},
  {Mulcahy}, {Munk}, {Nelles}, {Norden}, {Orru}, {Paas}, {Pandey-Pommier},
  {Pandey}, {Pietka}, {Pizzo}, {Polatidis}, {Reich}, {R{\"o}ttgering},
  {Scaife}, {Schwarz}, {Serylak}, {Smirnov}, {Stappers}, {Stewart}, {Tagger},
  {Tang}, {Tasse}, {Thoudam}, {Toribio}, {Vermeulen}, {van Weeren}, {Wucknitz},
  {Yatawatta}, \& {Zarka}}]{mo14}
{Morosan}, D.~E., {Gallagher}, P.~T., {Zucca}, P., {et~al.} 2014, \aap, 568,
  A67

\bibitem[{{Newkirk}(1961)}]{new61}
{Newkirk}, Jr., G. 1961, \apj, 133, 983

\bibitem[{{Saito} {et~al.}(1977){Saito}, {Poland}, \& {Munro}}]{sa77}
{Saito}, K., {Poland}, A.~I., \& {Munro}, R.~H. 1977, \solphys, 55, 121

\bibitem[{{Stappers} {et~al.}(2011){Stappers}, {Hessels}, {Alexov}, {Anderson},
  {Coenen}, {Hassall}, {Karastergiou}, {Kondratiev}, {Kramer}, {van Leeuwen},
  {Mol}, {Noutsos}, {Romein}, {Weltevrede}, {Fender}, {Wijers}, {B{\"a}hren},
  {Bell}, {Broderick}, {Daw}, {Dhillon}, {Eisl{\"o}ffel}, {Falcke},
  {Griessmeier}, {Law}, {Markoff}, {Miller-Jones}, {Scheers}, {Spreeuw},
  {Swinbank}, {Ter Veen}, {Wise}, {Wucknitz}, {Zarka}, {Anderson}, {Asgekar},
  {Avruch}, {Beck}, {Bennema}, {Bentum}, {Best}, {Bregman}, {Brentjens}, {van
  de Brink}, {Broekema}, {Brouw}, {Br{\"u}ggen}, {de Bruyn}, {Butcher},
  {Ciardi}, {Conway}, {Dettmar}, {van Duin}, {van Enst}, {Garrett}, {Gerbers},
  {Grit}, {Gunst}, {van Haarlem}, {Hamaker}, {Heald}, {Hoeft}, {Holties},
  {Horneffer}, {Koopmans}, {Kuper}, {Loose}, {Maat}, {McKay-Bukowski},
  {McKean}, {Miley}, {Morganti}, {Nijboer}, {Noordam}, {Norden}, {Olofsson},
  {Pandey-Pommier}, {Polatidis}, {Reich}, {R{\"o}ttgering}, {Schoenmakers},
  {Sluman}, {Smirnov}, {Steinmetz}, {Sterks}, {Tagger}, {Tang}, {Vermeulen},
  {Vermaas}, {Vogt}, {de Vos}, {Wijnholds}, {Yatawatta}, \& {Zensus}}]{sta11}
{Stappers}, B.~W., {Hessels}, J.~W.~T., {Alexov}, A., {et~al.} 2011, \aap, 530,
  A80

\bibitem[{{van Haarlem} {et~al.}(2013){van Haarlem}, {Wise}, {Gunst}, {Heald},
  {McKean}, {Hessels}, {de Bruyn}, {Nijboer}, {Swinbank}, {Fallows},
  {Brentjens}, {Nelles}, {Beck}, {Falcke}, {Fender}, {H{\"o}randel},
  {Koopmans}, {Mann}, {Miley}, {R{\"o}ttgering}, {Stappers}, {Wijers},
  {Zaroubi}, {van den Akker}, {Alexov}, {Anderson}, {Anderson}, {van Ardenne},
  {Arts}, {Asgekar}, {Avruch}, {Batejat}, {B{\"a}hren}, {Bell}, {Bell}, {van
  Bemmel}, {Bennema}, {Bentum}, {Bernardi}, {Best}, {B{\^i}rzan}, {Bonafede},
  {Boonstra}, {Braun}, {Bregman}, {Breitling}, {van de Brink}, {Broderick},
  {Broekema}, {Brouw}, {Br{\"u}ggen}, {Butcher}, {van Cappellen}, {Ciardi},
  {Coenen}, {Conway}, {Coolen}, {Corstanje}, {Damstra}, {Davies}, {Deller},
  {Dettmar}, {van Diepen}, {Dijkstra}, {Donker}, {Doorduin}, {Dromer}, {Drost},
  {van Duin}, {Eisl{\"o}ffel}, {van Enst}, {Ferrari}, {Frieswijk}, {Gankema},
  {Garrett}, {de Gasparin}, {Gerbers}, {de Geus}, {Grie{\ss}meier}, {Grit},
  {Gruppen}, {Hamaker}, {Hassall}, {Hoeft}, {Holties}, {Horneffer}, {van der
  Horst}, {van Houwelingen}, {Huijgen}, {Iacobelli}, {Intema}, {Jackson},
  {Jelic}, {de Jong}, {Kant}, {Karastergiou}, {Koers}, {Kollen}, {Kondratiev},
  {Kooistra}, {Koopman}, {Koster}, {Kuniyoshi}, {Kramer}, {Kuper},
  {Lambropoulos}, {Law}, {van Leeuwen}, {Lemaitre}, {Loose}, {Maat}, {Macario},
  {Markoff}, {Masters}, {McFadden}, {McKay-Bukowski}, {Meijering}, {Meulman},
  {Mevius}, {Millenaar}, {Miller-Jones}, {Mohan}, {Mol}, {Morawietz},
  {Morganti}, {Mulcahy}, {Mulder}, {Munk}, {Nieuwenhuis}, {van Nieuwpoort},
  {Noordam}, {Norden}, {Noutsos}, {Offringa}, {Olofsson}, {Omar}, {Orr{\'u}},
  {Overeem}, {Paas}, {Pandey-Pommier}, {Pandey}, {Pizzo}, {Polatidis},
  {Rafferty}, {Rawlings}, {Reich}, {de Reijer}, {Reitsma}, {Renting},
  {Riemers}, {Rol}, {Romein}, {Roosjen}, {Ruiter}, {Scaife}, {van der Schaaf},
  {Scheers}, {Schellart}, {Schoenmakers}, {Schoonderbeek}, {Serylak},
  {Shulevski}, {Sluman}, {Smirnov}, {Sobey}, {Spreeuw}, {Steinmetz}, {Sterks},
  {Stiepel}, {Stuurwold}, {Tagger}, {Tang}, {Tasse}, {Thomas}, {Thoudam},
  {Toribio}, {van der Tol}, {Usov}, {van Veelen}, {van der Veen}, {ter Veen},
  {Verbiest}, {Vermeulen}, {Vermaas}, {Vocks}, {Vogt}, {de Vos}, {van der Wal},
  {van Weeren}, {Weggemans}, {Weltevrede}, {White}, {Wijnholds}, {Wilhelmsson},
  {Wucknitz}, {Yatawatta}, {Zarka}, {Zensus}, \& {van Zwieten}}]{lofar13}
{van Haarlem}, M.~P., {Wise}, M.~W., {Gunst}, A.~W., {et~al.} 2013, \aap, 556,
  A2

\bibitem[{{Wentzel}(1984)}]{we84}
{Wentzel}, D.~G. 1984, \solphys, 90, 139

\bibitem[{{Zaitsev} \& {Zlotnik}(1986)}]{ze86}
{Zaitsev}, V.~V. \& {Zlotnik}, E.~Y. 1986, Soviet Astronomy Letters, 12, 128

\bibitem[{{Zucca} {et~al.}(2014){Zucca}, {Carley}, {Bloomfield}, \&
  {Gallagher}}]{zu14}
{Zucca}, P., {Carley}, E.~P., {Bloomfield}, D.~S., \& {Gallagher}, P.~T. 2014,
  \aap, 564, A47

\end{thebibliography}

\end{document}